\documentclass[preprint,aps,amssymb,superscriptaddress,nofootinbib]{revtex4}

\usepackage{graphics}
\usepackage{makeidx}
\usepackage{epsfig}
\usepackage{subfig}
\usepackage{colortbl}
\usepackage{colordvi}
\usepackage{verbatim}
\usepackage{amsmath, amsthm}
\usepackage{enumerate}
\usepackage{wrapfig}
\usepackage{hyperref}
\usepackage{dcolumn}
\usepackage{bm}
\usepackage{bbm}
\usepackage{amssymb,mathrsfs}

\begin{document}
 
\title{How is Lorentz Invariance encoded in the Hamiltonian?}

\author{Nirmalya Kajuri}
\email{nirmalya@imsc.res.in}
 \affiliation{ Department of Physics, Indian Institute of Technology Madras,\\ Chennai 600036, India}

\begin{abstract} One of the  disadvantages of the Hamiltonian formulation is that Lorentz invariance is not manifest in the former. Given a Hamiltonian, there is no simple way to check whether it is relativistic or not. One would either have to solve for the equations of motion or calculate the Poisson Brackets of the Noether charges to perform such a check. In this paper we show that, for a class of Hamiltonians, it is possible to check Lorentz invariance directly from the Hamiltonian. Our work is particularly useful for theories where the other methods may not be readily available. 
\end{abstract}

\maketitle
\section{Introduction}
Consider the following Hamiltonian: 

 \begin{align} \label{hmail}
H = \frac{1}{2} ( \pi_A^2 + (\vec{\nabla}A)^2 +m_A^2 A^2 +\pi_B^2  +(\vec{\nabla}B)^2 + m_B^2B^2)+ \frac{g}{4} (\vec{\nabla}A)^2 B^2 + \frac{1}{2} \pi_A^2 \left(\frac{\frac{g}{2}B^2}{1+\frac{g}{2}B^2}\right)
\end{align}

where $A$, $B$ are scalar fields and $g$ is a coupling constant. Is Lorentz invariance a symmetry of the system described by the above Hamiltonian? The answer is yes. The corresponding Lagrangian is given by:
\begin{align}\label{lag} 
L = \frac{1}{2} ( (\partial_{\mu} A)^2 - m_A^2 A^2 +(\partial_{\mu} B)^2 -m_B^2B^2)-\frac{g}{4}(\partial_{\mu} A)^2 B^2  
\end{align}

Which is manifestly Lorentz invariant. But this was not at all obvious from the Hamiltonian. Given a Hamiltonian how do we check for Lorentz invariance?

Generally, one has three avenues :
One, solve for the equations of motion and see if these are relativistic. 

Two, calculate the Poisson brackets of the Noether charges of Lorentz symmetry with each other and check if this gives a representation of the Lorentz algebra. We do not elaborate on this method here, the reader is referred to \cite{greiner} for details. The non-trivial part of this method involves calculating the Poisson Bracket relations of the generators of boost and angular momenta with the Hamiltonian.

Finally, one may construct the Lagrangian and check if it is Lorentz invariant. If one is given a quantum Hamiltonian, then this last method would involve obtaining a path integral representation. 

In this paper we will show that for a class of physical systems, it is possible to obtain the condition for Lorentz invariance directly from a  Hamiltonian. As we will see, the condition states that certain vectors and tensors that may be constructed from the Hamiltonian are proportional to one another. We note that this method works when the representation of Lorentz symmetry is linear. There are cases  where Lorentz symmetry is non-linearly realized, for instance gauge-fixed or reduced phase space treatments of gauge theories (see for instance \cite{Bernstein:1974rd}), but these are not considered here.

The systems for which the result holds must satisfy certain criteria. These are: 
where the field momentum or the space derivative of one of the fields occurs in the Lagrangian they must 

(i)  Come in even powers . That is, any of the terms in the Lagrangian may contain a term like $(\partial_{x} A)^2 $ but a factor like $(\partial_{\mu} A) C^{\mu}$, where $C^{\mu}$ may be a (pseudo)vector, cannot occur in any of the terms.

(ii) Not be multiplied with the derivatives of one of the other fields. That is a term like $(\partial_{x} A)^2 B^2$ is allowed but $(\partial_{x} A)^2(\partial_{x} B)^2$ is not. 

This work is particularly important in the context of those quantum theories where the other avenues may not be readily available. One such case was polymer quantized scalar field theory \cite{Ashtekar:2002vh}. The status of Lorentz invariance of this theory was open till a path integral formulation was found recently \cite{Kajuri:2014kva}, establishing that the theory is not Lorentz invariant. However our work makes the deduction of Lorentz non-invariance of this theory extremely simple, as we will show. 

The plan of the paper is as follows. In the next section we start from the Lagrangian formulation and show how to derive the condition for Lorentz invariance for these Hamiltonians. First we restrict ourselves to an even more limited (but still non-trivial) class of Hamiltonians, for which the condition for Lorentz invariance takes a particularly simple form.  Then after presenting some examples we sketch how to extend the derivation for all Hamiltonians satisfying (i) and (ii) above. The final section summarizes our findings. 
\section{Encoding Relativistic Invariance in the Hamiltonian: The Restricted Case}
In this section we derive a sufficient condition for Lorentz invariance for physical systems which satisfy (i) and (ii) above. First we will present the derivation for systems which satisfy another criteria : 

(iii)  The derivative terms appearing in the Lagrangian must be quadratic. That is, any of the terms in the Lagrangian may contain the factor $(\partial_{x} A)^2 $ but a factor like $(\partial_{\mu} A)^4$  cannot occur in any of the terms. 

  Note that the system described by \eqref{lag} (equivalently \eqref{hmail}) satisfies these criteria.

For all such Hamiltonians, the condition for Lorentz invariance takes a particularly simple form. It goes as: construct for each field A, the following two column matrices - 
\begin{align} 
\label{def1}F_A^{\mu}&=(\frac{\partial H}{\partial \pi_A}, \vec{\nabla}A) \\
\label{def2}G_A^{\mu} &=( \pi_A, \frac{\partial H}{\partial  \vec{\nabla}A})
\end{align}
  The condition for Lorentz invariance is, for each $A$:

\begin{align}\label{condition}
F_A^{\mu} = k G_A^{\mu}
\end{align}
where $k$ is a scalar which may be the function of fields and its derivatives.  

In the following subsection we derive \eqref{condition} starting from a Lagrangian formulation. This will be followed by some examples. We will complete this section by sketching the steps of the derivation for a system satisfying only (i) and (ii). 
 
\subsection{Derivation from the Lagrangian}
In this section we will demonstrate how, for systems satisfying criteria (i) and (ii), the Lorentz invariance of a Lagrangian is encoded in the Hamiltonian through \eqref{condition}. We'll exhibit the proof for a system of scalar fields. The extension to higher spin fields is straightforward. 

Let us consider a relativistic Lagrangian L which is a function of scalar fields $\phi_n$. We write this as:
$$L = L_1 +L_2 $$ 
where all terms containing derivatives in the field are put in $L_1$ and the rest of the terms are in $L_2$. For instance, in the Lagrangian \eqref{lag} we will have 
\begin{align}
L_1 = \frac{1}{2} ( (\partial_{\mu} A)^2   +(\partial_{\mu} B)^2 )+\frac{g}{4}(\partial_{\mu} A)^2 B^2  
L_1 = - m_A^2 A^2  -m_B^2B^2
\end{align}

Now since $L_2$ does not contain any terms containing field derivatives, it follows that $L_1$ and $L_2$ must be separately Lorentz invariant. Now for a Lagrangian satisfying the criteria (i) and (ii), $L_1$ may be written as: 
\begin{align}\label{lag1}
L_1 = \sum_n \frac{1}{2} \frac{\partial L_1} {\partial (\partial_{\mu} \phi_n)} \partial_{\mu} \phi_n
\end{align}
Again, each term in $L_1$ must be Lorentz invariant in itself. But from the above, each term in  $L_1$ can be written in the form 
\begin{align}
(L_1)_n =\frac{1}{2} \eta_{\mu \nu} W_n^{\mu} Z_n^{\nu}
\end{align}
where
\begin{align}
W_n^{\mu} =  \partial_{\mu} \phi_n\\
Z_n^{\nu} = \frac{\partial L_1} {\partial (\partial_{\mu} \phi_n)}
\end{align}

This is manifestly a Lorentz invariant quantity if $W_n^{\mu}$ and $Z_n{\nu}$ transform as four vectors under Lorentz transformations. But $W_n^{\mu}$ are a Lorentz four vectors by definition. It therefore follows that the condition of Lorentz invariance of the Lagrangian $L$ translates to the condition that the terms $Z_n{\nu}$ transform as four vectors under Lorentz transformations. Now for a given n  $Z_n^{\mu}$ is formed by differentiating the Lagrangian with respect to spacetime derivatives of the field $\phi_n$. Therefore $Z_n^{\mu}$ must contain spacetime derivatives of the field itself. Therefore to transform similarly as $W_n^{\mu}$, it must be that 
\begin{align}\label{condi}
W_n^{\mu} = kZ_n^{\mu}
\end{align}
where k is a scalar which may be a function of the fields and their derivatives. This relation therefore encodes the Lorentz invariance of the Lagrangian. 

Now let us proceed to construct the Hamiltonian:
\begin{align}
\nonumber H &= \sum_n \frac{\partial L} {\partial (\partial_{0} \phi_n)} \partial_{0} \phi_n - L_1 -L_2 \\
&= \sum_n \left(\frac{\partial L_1} {\partial (\partial_{0} \phi_n)} \partial_{0} \phi_n -\frac{1}{2} \frac{\partial L_1} {\partial (\partial_{\mu} \phi_n)} \partial_{\mu} \phi_n \right)-L_2\\
&= -L_2 + \sum_n \frac{1}{2} \left(\frac{\partial L_1} {\partial (\partial_{0} \phi_n)} \partial_{0} \phi_n + \frac{\partial L_1} {\partial (\partial_{i} \phi_n)} \partial_{i} \phi_n \right)\\
&= -L_2 + \sum_n \frac{1}{2}\delta_{\mu \nu} W_n^{\mu} Z_n^{\nu}
\end{align}
As $\eta_{\mu \nu} W_n^{\mu} Z_n^{\nu}$ was invariant under Lorentz transformations, it follows that $\delta_{\mu \nu} W_n^{\mu} Z_n^{\nu}$ would be invariant under orthogonal transformations.
  Let us convert these into functions of $\pi_n$ and $\phi_n$. We define
\begin{align}
&F_n^0 \equiv  W_n^{0} =  \partial_{0} \phi_n = \frac{\partial_{0} H}{\partial \pi_n} \\
&F_n^i \equiv  W_n^{i} = \partial_{i} \phi_n \\
&G_n^0 \equiv Z_n^{0} = \frac{\partial L_1} {\partial (\partial_{0} \phi_n)} =\pi_n\\
&G_n^i \equiv Z_n^{i} = \frac{\partial L_1} {\partial (\partial_{i} \phi_n)} =\frac{\partial H} {\partial (\partial_{i} \phi_n)}
\end{align}
Where we have used one of the Hamilton's equations of motion in the first step.  
Now if we decompose the Hamiltonian into $H_1$ and $H_2$  using the same logic as we used for the Lagrangian, we will have $H_2 = L_2$ and the Hamiltonian may be written as
\begin{align}
H =  H_2 + \sum_n  \frac{1}{2}\delta_{\mu \nu} F_n^{\mu} G_n^{\nu}
\end{align}
 
The condition \eqref{condi}  for Lorentz invariance of the Lagrangian becomes, in terms of phase space functions, exactly the condition \eqref{condition} advertised before: 
 \begin{align}\label{condi2}
F_n^{\mu} = k G_n^{\mu}
\end{align}
Thus for the given class of systems, the relativistic invariance of the dynamics can be read off from the Hamiltonian by constructing the column matrices $F_n^{\mu}$ and $G_n^{\mu}$ and checking if \eqref{condi2} is satisfied. 
 
\subsection{ A couple of examples}
As a first example let us consider the Hamiltonian of \eqref{hmail}. We know from \eqref{lag} that this is Lorentz invariant. Now let us check whether it satisfies our criteria. 
For the A field the relevant vectors are 
\begin{align}
F_n^{\mu}= (\frac{\partial H}{\partial \pi_A}, \vec{\nabla}A) = \left( \left( \frac{1+gB^2}{1+\frac{g}{2}B^2}\right) \pi_A, \vec{\nabla}A \right)\\
G_n^{\mu} =( \pi_A, \frac{\partial H}{\partial  \vec{\nabla}A}) = \left(\pi_A, \left( 1+\frac{g}{2}B^2\right)\vec{\nabla}A \right)
\end{align}
It is easy to check that for this case
 \begin{align}  
F_n^{\mu} = k G_n^{\mu}
\end{align}
with 
$$k  = \left( 1+ \frac{g}{2}B^2 \right)^{-1}$$

Let us consider another example, this time from polymer quantization of scalar fields \cite{Ashtekar:2002vh}. That this system is not Lorentz invariant is known from its path integral formulation \cite{Kajuri:2014kva}. Here we have the following Hamiltonian:
 \begin{align}
H = \frac{1}{2} \left( \pi^2 + (\nabla \phi)^2 \cos^2 (\mu \phi)\right)
\end{align} 
where $\mu$ is a dimensionless quantity.
Let us consider the vectors for this case: 
 \begin{align}
F_n^{\mu} =( \pi, \nabla \phi)\\
G_n^{\mu} =(\pi, \nabla \phi)\cos^2 (\mu \phi))
\end{align} 

Clearly these are not proportional to each other. Thus our method agrees with the known result. 
\subsection{Encoding Relativistic Invariance in the Hamiltonian: more general case}
In this section we briefly sketch how to obtain the criteria for relativistic invariance for systems satisfying the criteria (i) and (ii) only.

The Lagrangian for such a system may be divided into $L_1$ and $L_2$ as before and $L_1$ may be expanded as:
\begin{align}
L_1 = \frac{1}{2} \frac{\partial L_1}{\partial(\partial_{\mu} \phi)} \partial_{\mu} \phi + \frac{1}{4} \frac{ \partial^2 L_1}{\partial(\partial_{\mu} \phi) \partial( \partial_{\nu}\phi)}   \partial_{\nu}\phi \partial_{\mu} \phi +...
\end{align}

This may be written as 
\begin{align}
L_1 = \frac{1}{2} \eta_{\mu \nu} A^{\mu} B^{\nu} + \frac{1}{4}\eta_{\alpha \beta} \eta_{\gamma \chi} C^{\alpha  \gamma} B^{\beta} B^{\chi} +....
\end{align}

where 
\begin{align}
A^{\mu}= \frac{\partial L_1}{\partial(\partial_{\mu} \phi)}\\
 B^{\mu} =  \partial_{\mu} \phi\\
C^{\mu \nu} =  \frac{ \partial^2 L_1}{\partial(\partial_{\mu} \phi) \partial( \partial_{\nu}\phi)} 
\end{align}
and so on.
Again this expression is manifestly Lorentz invariant, given that $A^{\mu} \propto B^{\mu},C^{\mu \nu}\propto B^{\mu}B^{\nu}...$ by the same logic as before. Now all these terms may be expressed in the Hamiltonian formulation in terms of H, $\pi, \vec{\nabla} \phi$ and derivatives of H to obtain the conditions in the Hamiltonian language, just as before. 

\section{Summary and Outlook}
In this paper we have shown that, for a class of systems, there exists a simple way to check for Lorentz invariance directly from the Hamiltonian formulation. This class of systems were defined by conditions (i) and (ii) given above. For an even more restricted class, defined by conditions (i), (ii) \textit{and} (iii) given in section II, the condition reduces to a simple proportionality between two column matrices which may be constructed from the Hamiltonian.

This is particularly important in the context of quantum theories where other methods of checking for Lorentz invariance may be immediately available. One place where this is the case is polymer quantized field theory. Here the only check on Lorentz invariance so far has come from the path integral formulation \cite{Kajuri:2014kva}. We managed to reach the same conclusion in a much simpler way here.

\begin{acknowledgements}
 We thank James Edwards for illuminating discussions and helpful comments and Gaurav Narain for helpful comments on the draft.
\end{acknowledgements}

\end{document}